\documentclass[11pt]{article}
\usepackage{times,newpasp,epsf}
\def\etal{et~al.}

\def\msun { \rm {M_\odot}}

\def\t0{t_{\rm 0}}

\def\spose#1{\hbox to 0pt{#1\hss}}
\def\simlt{\mathrel{\spose{\lower 3pt\hbox{$\mathchar"218$}}
     \raise 2.0pt\hbox{$\mathchar"13C$}}}
\def\simgt{\mathrel{\spose{\lower 3pt\hbox{$\mathchar"218$}}
     \raise 2.0pt\hbox{$\mathchar"13E$}}}

\def\that {\widehat{t}}

\def\edcomment#1{\iffalse\marginpar{\raggedright\sl#1\/}\else\relax\fi}
\reversemarginpar

\begin{document}
\title{\large{The} \LARGE{G}\large{alactic} \LARGE{E}\large{xoplanet} 
\LARGE{S}\large{urvey} \LARGE{T}\large{elescope}: \\
A Proposed Space-Based Microlensing Survey for Terrestrial
Extra-Solar Planets}
 \author{David P.~Bennett \& Sun Hong Rhie}
\affil{Department of Physics, University of Notre Dame,
                 Notre Dame, IN 46556}

\begin{abstract}
We present a conceptual design for a space based Galactic Exoplanet
Survey Telescope (GEST) which will use the gravitational microlensing
technique to detect extra solar planets with
masses as low as that of Mars at all separations $\simgt 1\,$AU.
The microlensing data would be collected
by a diffraction limited, wide field imaging telescope of
$\sim 1.5\,$m aperture equipped with a large array of red-optimized CCD
detectors. Such a system would be able to monitor $\sim 2\times 10^8$
stars in $\sim 6$ square degrees of the Galactic bulge at intervals of
20-30 minutes, and it would observe $\sim 12000$ microlensing
events in three bulge seasons. If planetary systems like our own are common,
GEST should be able to detect
$\sim 5000$ planets over a 2.5 year lifetime. If gas giants like Jupiter and
Saturn are rare, then GEST would detect $\sim 1300$ planets in a 2.5 year
mission if we assume that most planetary systems are dominated by
planets of about Neptune's' mass. Such a mission would also discover 
$\sim 100$ planets of an Earth mass or smaller if such planets are common.
This is a factor of $\sim 50$ better than the most ambitious ground based
programs that have been proposed. GEST will also be sensitive to planets
which have been separated from their parent stars.
\end{abstract}

\section{Microlensing Planet Detection: GEST's Main Goal}
\label{sec-intro}
 
The main strength of the gravitational microlensing planet search
technique is that it is sensitive to lower mass planets than other
techniques. Observed from space, signals for planets down to the mass of
Mars are
detectable, but they are much rarer and have a shorter duration than higher
mass planetary signals. Thus, a large number of stars must be followed
with a high sampling frequency in order to detect low mass planets. With
GEST, we are able to monitor $\sim 2 \times 10^8$ stars once every 30 
minutes or so with a photometric accuracy of $\sim 1\,$\%. 
The microlensing event will not repeat, so high quality photometric
data must be obtained while the event is in progress. 
Our proposed GEST mission will accomplish this.
 
The planetary systems studied by the microlensing technique are located
$1-8\,$kpc away towards the Galactic center rather than in 
the local neighborhood. The planetary signals
will usually be detected as a  modifications of the single lens light 
curve due to the gravitational effect of the planet.
Microlensing is most sensitive to planets near the Einstein ring radius
which corresponds to a distance of $1-10\,$AU from the lens star, but
because GEST does not require the discovery of stellar microlensing event
to begin intensive monitoring for planets, GEST will be able to detect
planets at arbitrarily large separations from their host stars. One such
isolated planet may have already been observed (Bennett et al 1997)
by the MACHO Collaboration.

\section{Required Features of the Satellite Design}
\label{sec-satfeat}
 
The main challenge for a microlensing search for low mass planets is that,
although the planetary signals can be strong, they are both very rare and
and have durations as short as $\sim 2\,$hours
(Mao \& Paczynski 1991; Gould \& Loeb 1992). Furthermore, the relatively
large angular size of giant source stars makes them poor targets for a low
mass microlensing planet search project as finite source effects (Bennett \&
Rhie 1996) tend to wash out the microlensing signal.
Thus, GEST must be able to monitor 
large areas of sky rapidly while attaining $\sim 1$\% photometry on main
sequence source stars. In the central Galactic bulge fields where the 
microlensing rate is highest, ground based images are seriously incomplete
at or above the bulge main sequence turn-off, so there is a great advantage
to be gained from higher resolution imaging from space which will allow
many of the main sequence stars to be separately resolved.
Since areas of high star density
must be observed, we need high angular resolution to maintain
photometric accuracy. So, we are led to a requirement of a large
field of view and a high angular resolution--for a very large number of
CCD pixels. Our baseline design calls for a $\sim 1$ square degree field of
view with pixels of $0.1$" or smaller, for a total of $\simgt 1.3\times 10^9$
pixels. Our suggested observing strategy will be to cycle over 4-6
selected fields near the Galactic center every 20-30 minutes. Given the
high data rate implied by this strategy, a geosynchronous orbit seems
sensible.
 
Our selected Galactic bulge fields fields have high reddening, so it
is advantageous to observe at long wavelengths. Fortunately, recent
developments in CCD technology have produced devices with a quantum
efficiency of $\simgt 80$\% at 750-950nm (Groom et al 2000).
This is $\sim 4$ times better than
the CCD's used by HST's WFPC2 camera.

\section{A Simulation of the GEST Mission}
\label{sec-sim}
 
\begin{figure}
\plottwo{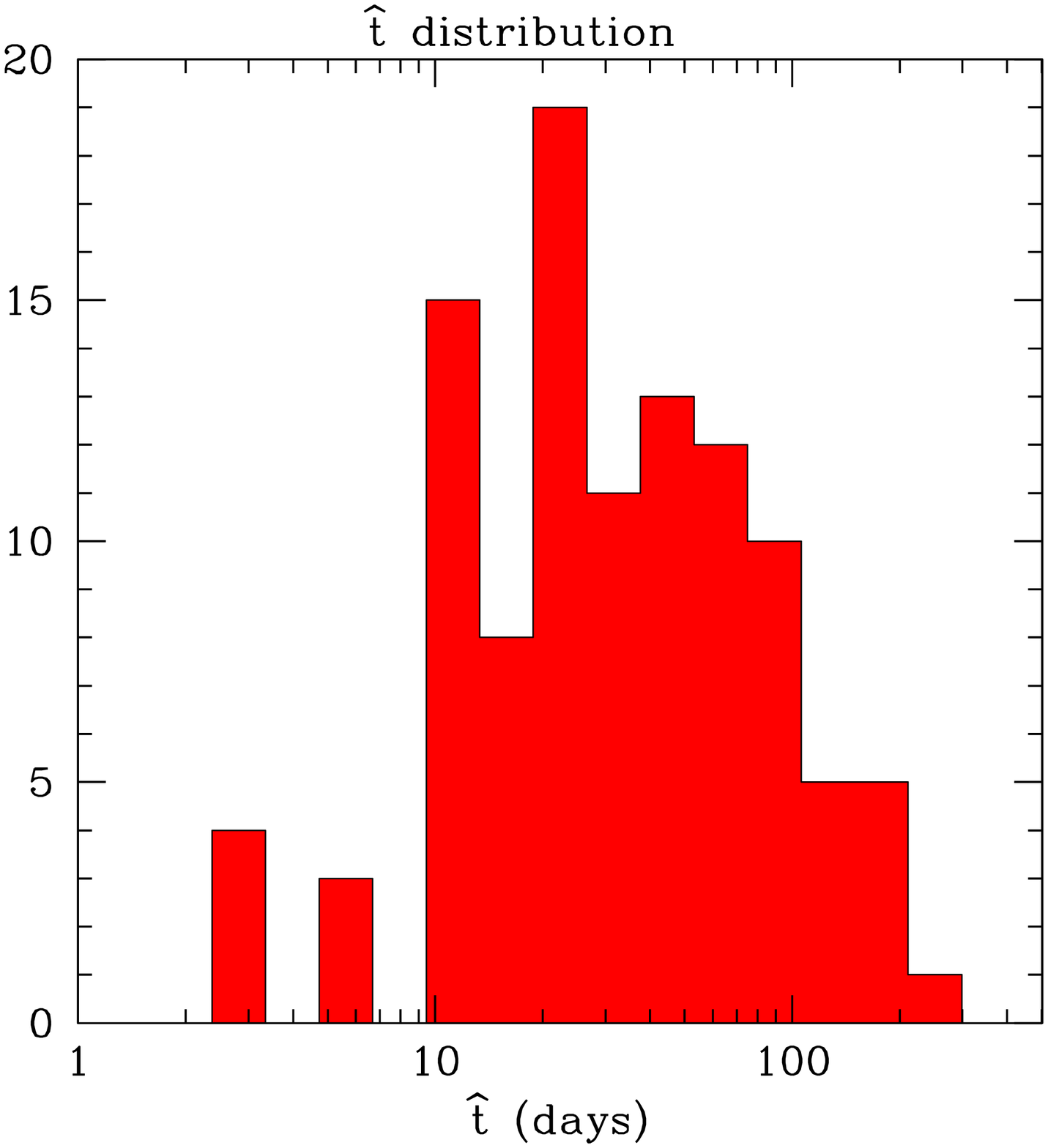}{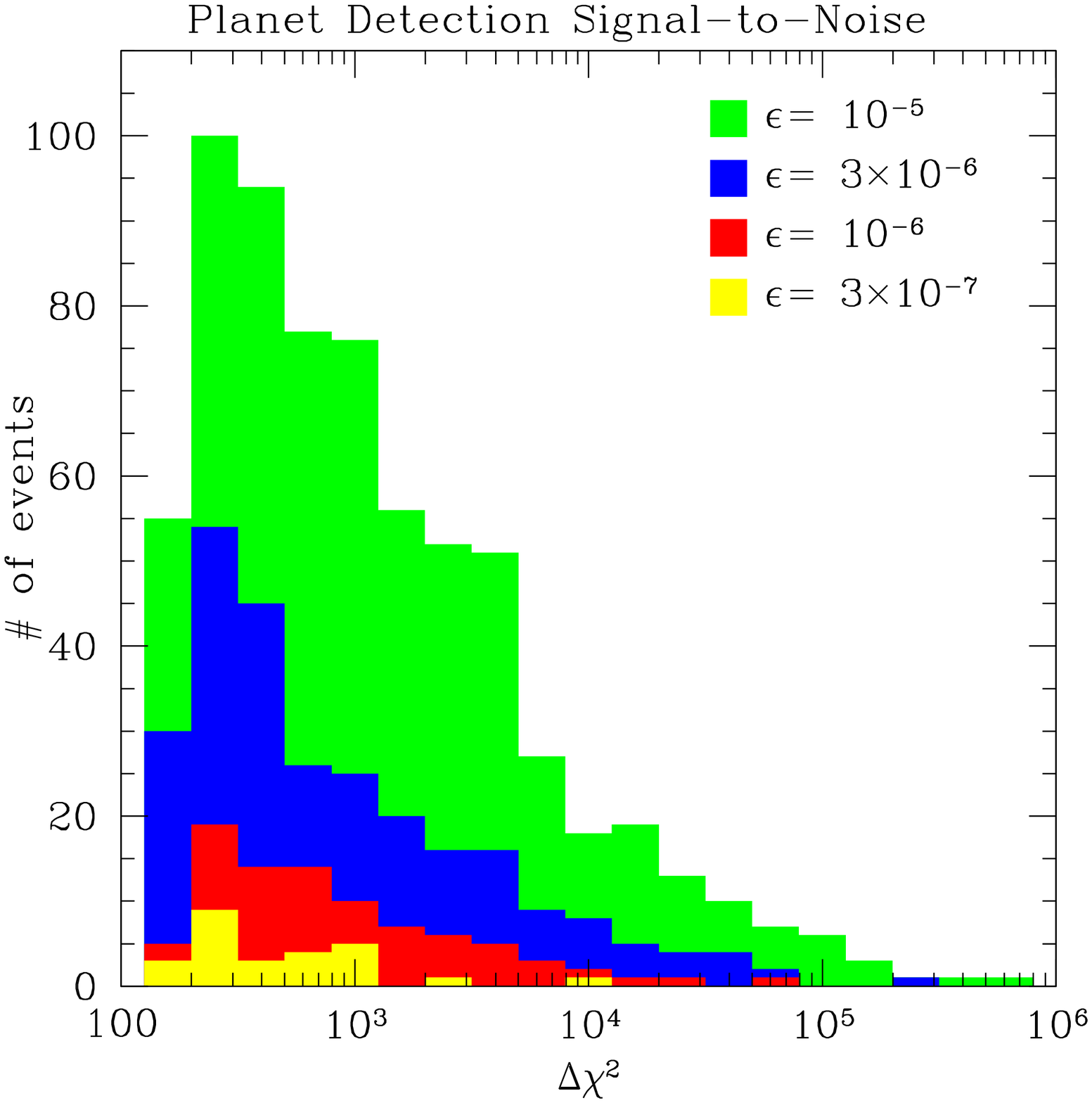}
\caption{
The distribution of Einstein diameter crossing times ($\that$), corrected
for detection efficiencies, is shown in the left hand panel, 
and the planetary detection signal-to-noise distribution is shown on the
right as a function of the planetary mass fraction $\epsilon$.
\label{fig}}
\end{figure}
 
We have carried out a detailed simulation of the expected
performance of the GEST mission. Here are the relevant features
of our simulation: We assume that 6 square degrees of the Galactic bulge are
observed once every 27 minutes. Photometry is assumed to be done using 
a difference imaging technique which is photon noise limited with a 0.3\%
systematic error added in quadrature. The photon 
noise is due to the target star and also any neighboring stars which
have a PSF that overlaps with the target star. The effect of
the neighboring stars was taken into account by constructing
an artificial image with a resolution of 0.16" (the diffraction
limit for a 1.5m telescope at $\lambda = 900\,$nm).
We assume a signal to noise of 60 for a source with $I = 22$, or
about 16 detected photo-electrons per second for a 225 sec exposure. 
This estimate assumes the use of high sensitivity CCD's and a 
broad passband of 750-950nm.

We assume that the luminosity function found by Holtzman \etal\ (1998)
for Baade's window applies to the entire 6 square degrees. This
is conservative in that most of our fields will be closer to the
Galactic Center with a higher star density, but this will be
partially compensated for by the higher reddening in these fields.
We conservatively assume a microlensing optical depth of 
$\tau = 3\times 10^{-6}$ which is slightly below the measured values
(Udalski et al, 1994; Alcock et al 1997),
and we assume the event timescale
distribution given by Alcock \etal\ (2000) as shown in Figure 1.
 
\section{GEST Simulation Results}
\label{sec-results}

\begin{figure}
\plottwo{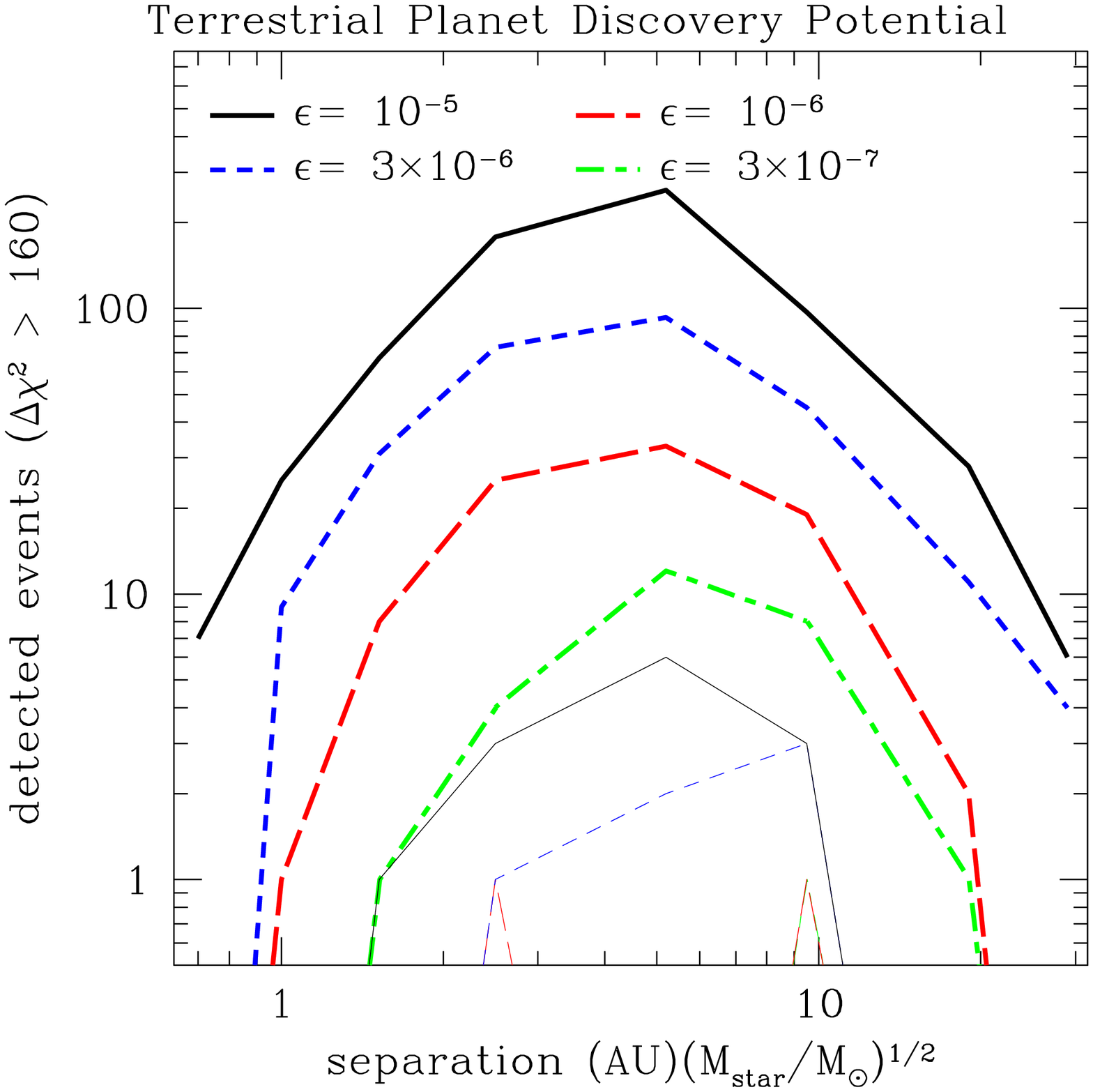}{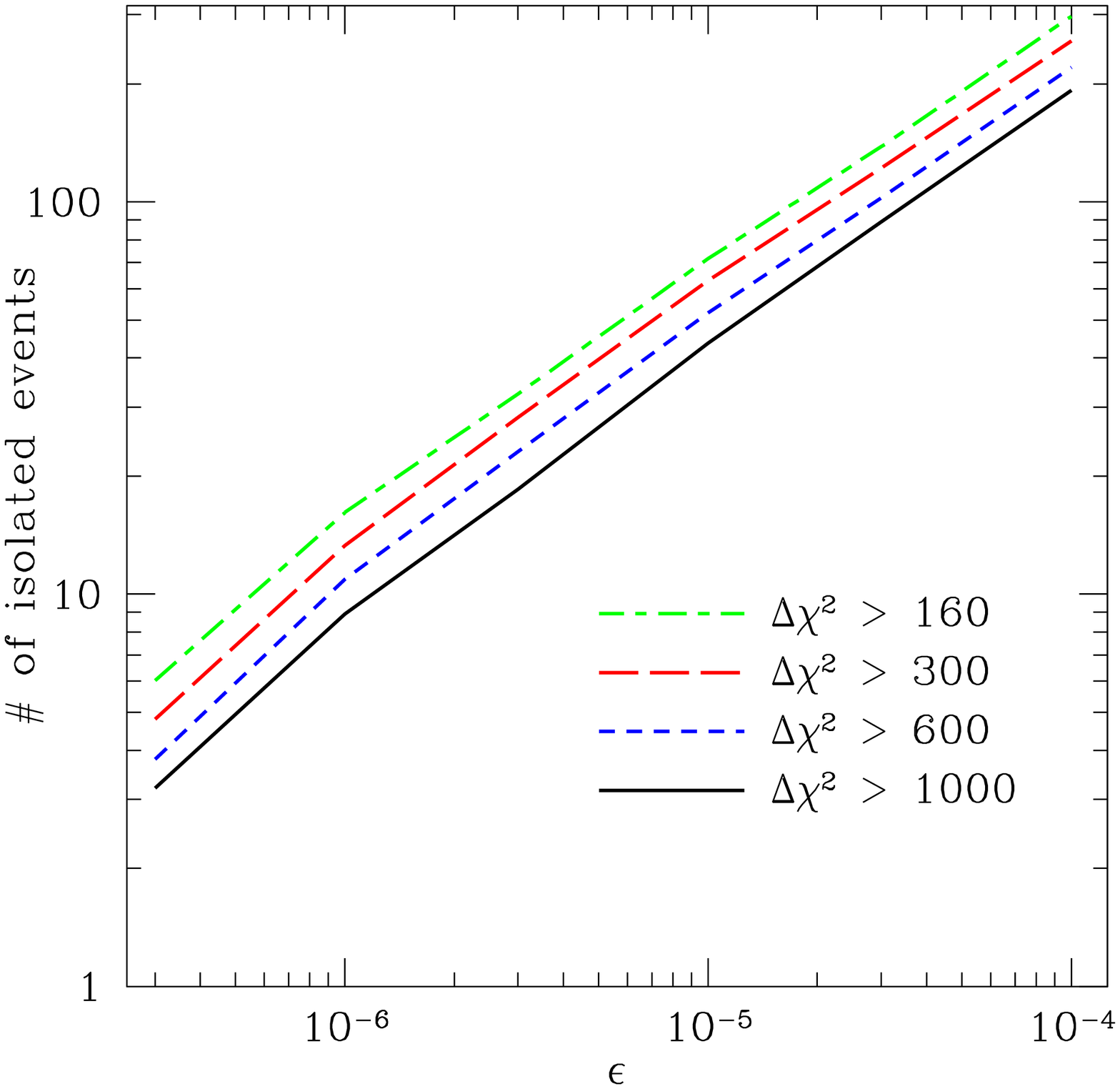}
\caption{
The left panel shows the number of potential planetary detections 
plotted as a function
of planetary mass fraction and separation for a selection criteria of
$\Delta\chi^2 \geq 160$ for the GEST (upper thicker lines), and the VST
(lower lines) surveys. This plot is normalized to the expected number 
of GEST planetary discoveries assuming every planetary system has a planet 
of the given mass fraction at the given separation. The right panel shows
the number of isolated planet detections as a function of the planetary 
mass fraction ($\epsilon$) under the assumption that there is one isolated
planet for every star in the Galaxy.
\label{fig-gestlgn160}}
\end{figure}

\begin{figure}
\plottwo{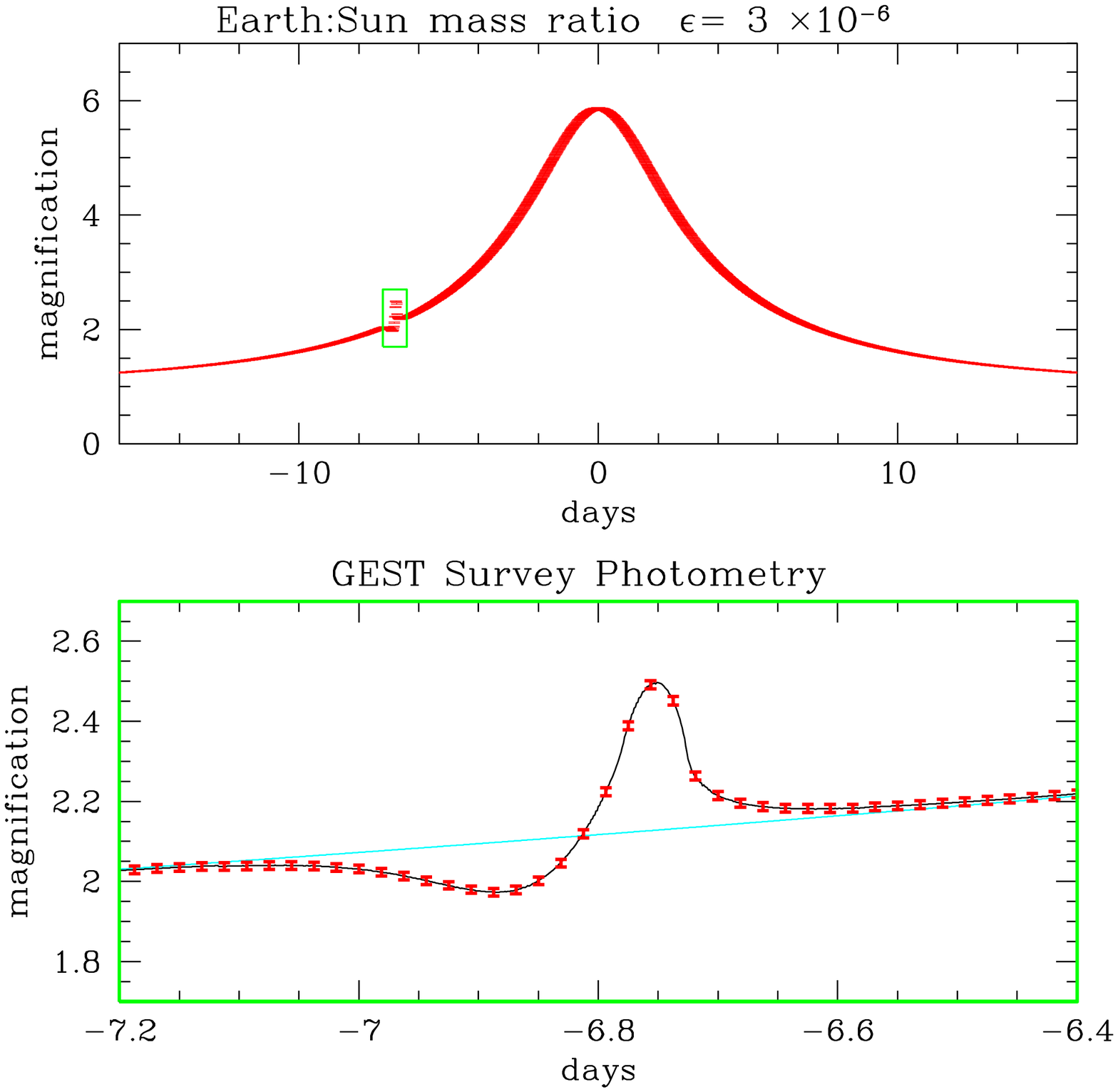}{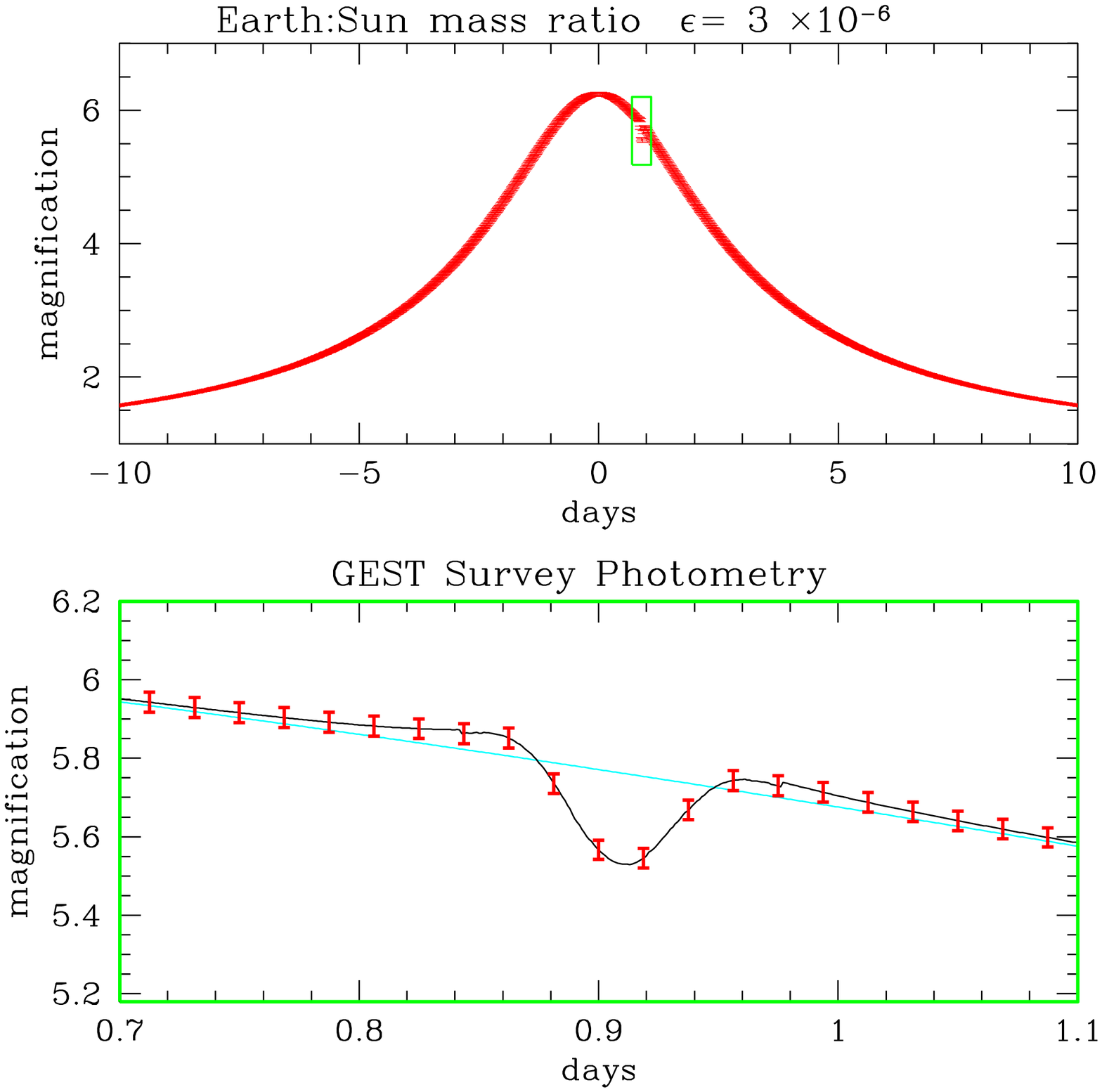}
\caption{
The simulated data for two Earth mass ratio (planetary mass
fraction $\epsilon = 3\times 10^{-6}$) planets with planet-star
separations of $a = 0.77$ and $a= 0.93$ Einstein ring radii, respectively. 
The top panels show the
simulated data for each event, and the region of the planetary deviation
is indicated with the green box. The lower panel shows a blow-up of the
planetary deviation region along with the planetary lensing lightcurve
and the best fit standard single lens lightcurve (in cyan).
The lightcurve on the left is a high signal-to-noise detection with
$\Delta \chi^2 = 4100$ while the lightcurve on the right is a detection
just above our signal-to-noise cut at $\Delta \chi^2 > 160$.
\label{fig-lc1}}
\end{figure}

\begin{figure}
\plottwo{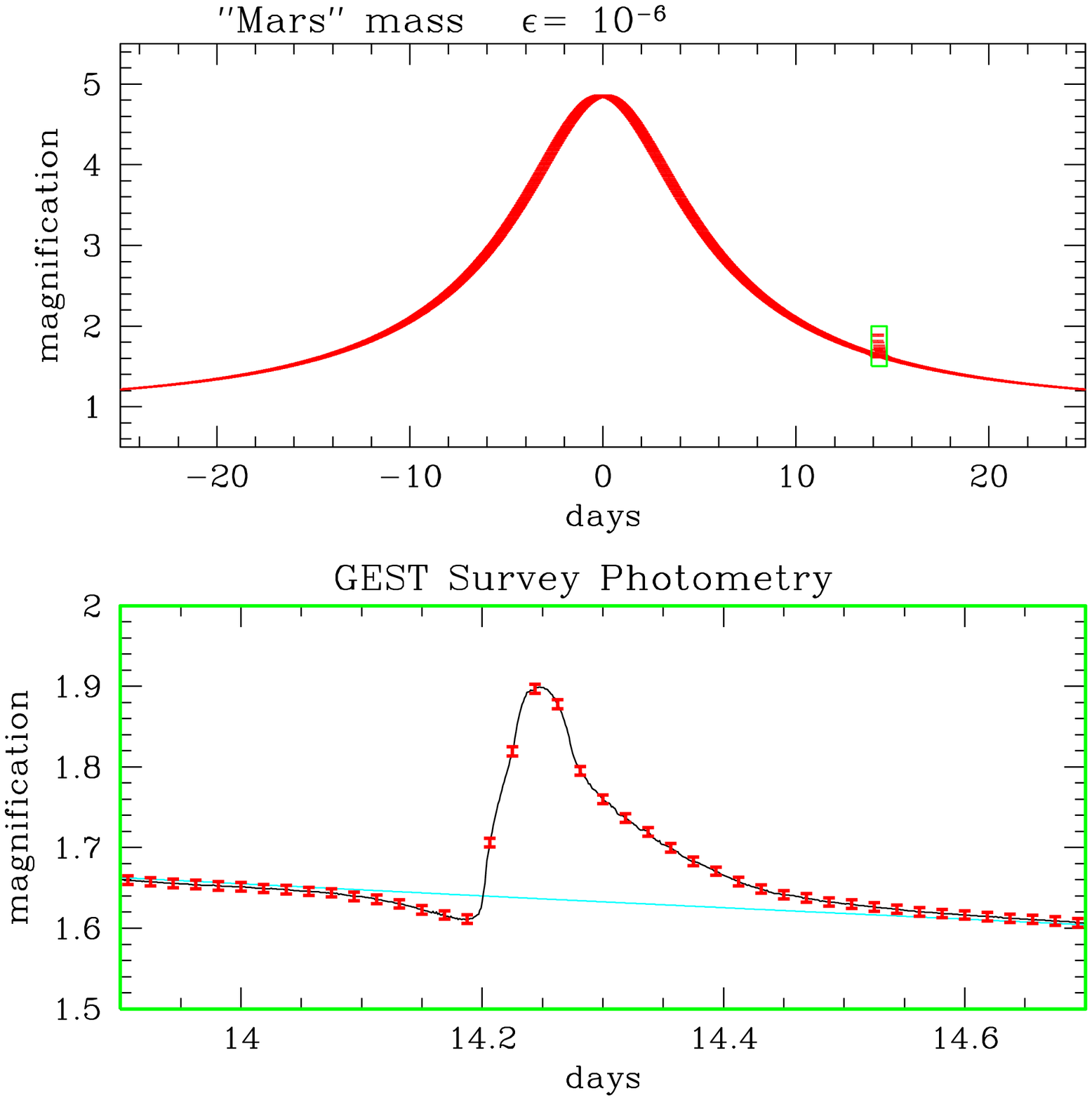}{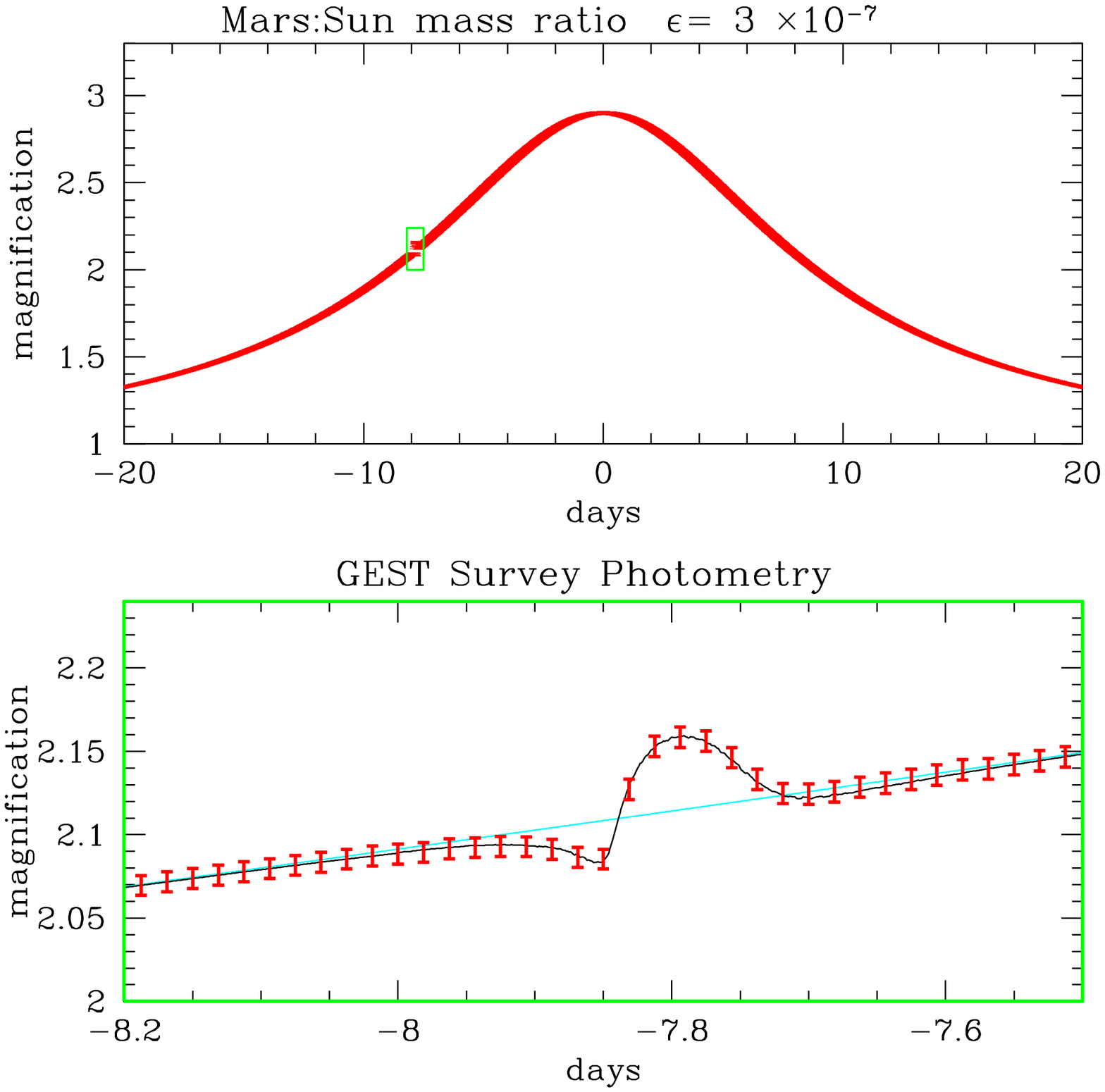}
\caption{
Simulated data for two planetary microlensing events. The event on the left
has a mass fraction
of $\epsilon = 10^{-6}$ and a separation of $a = 1.43$, and
the event on the right has $\epsilon = 3\times 10^{-7}$ and $a = 1.30$.
These represent a Mars mass planet (for a typical $M_{\rm star}\sim 0.3\msun$)
and a Mars mass ratio planet, respectively. These events 
have $\Delta \chi^2 = 7800$ and $\Delta \chi^2 = 200$, respectively.
\label{fig-lc6}}
\end{figure}
 
Our simulated GEST mission observes a total of 18,000 microlensing events
with peak magnification $> 1.34$ for source stars down to $I = 24.5$
over an assumed 3 year mission which observes the Galactic bulge for
8 months per year. Our detection threshold is a cut on $\Delta\chi^2$ which
is the $\chi^2$ difference between a single lens and planetary binary lens fit.
We've simulated planetary microlensing lightcurves for
planets at a range of separations ranging from $0.7-30(M_{\rm star}/\msun)$AU, 
and the left hand panel of Figure 2 shows the number of detected planets
if each lens system has a planet of the assumed mass fraction ($\epsilon$) at
the assumed separation. The right hand panel of Figure 2 shows the number of
detected ``isolated" planets as a function of the planetary mass fraction,
$\epsilon$, which is now compared to the average stellar mass.
Figure 2 indicates that we'll be able to detect 50-100 Earth mass planets
if they are common at distances of a few AU or more, and 10-20 Earth mass
planets if they are common at 1 AU. We also expect to detect $\sim$ 10 Mars
mass planets if they are common at a few AU. The right panel of Figure 1
shows the distribution of low mass planet detections as a function of
our signal-to-noise parameter $\Delta\chi^2$, and it indicates that most of
the detections are well above our nominal detection threshold of 
$\Delta\chi^2 = 160$. Examples of simulated GEST low mass planetary lightcurves
are shown in Figures 3-7.

For more massive planets, our sensitivity is much better. For solar system
analogs, our simulations indicate that GEST would detect $\sim 5000$ gas giant
planets if our solar system is typical. If the typical planetary system
resembles the solar system with Saturn and Jupiter replaced by Neptune mass
planets, then GEST would detect $\sim 1300$ of these Neptune-like planets.
 
\section{Comparison to Ground Based Surveys and Conclusions}
\label{sec-sample}
 
\begin{figure}
\plottwo{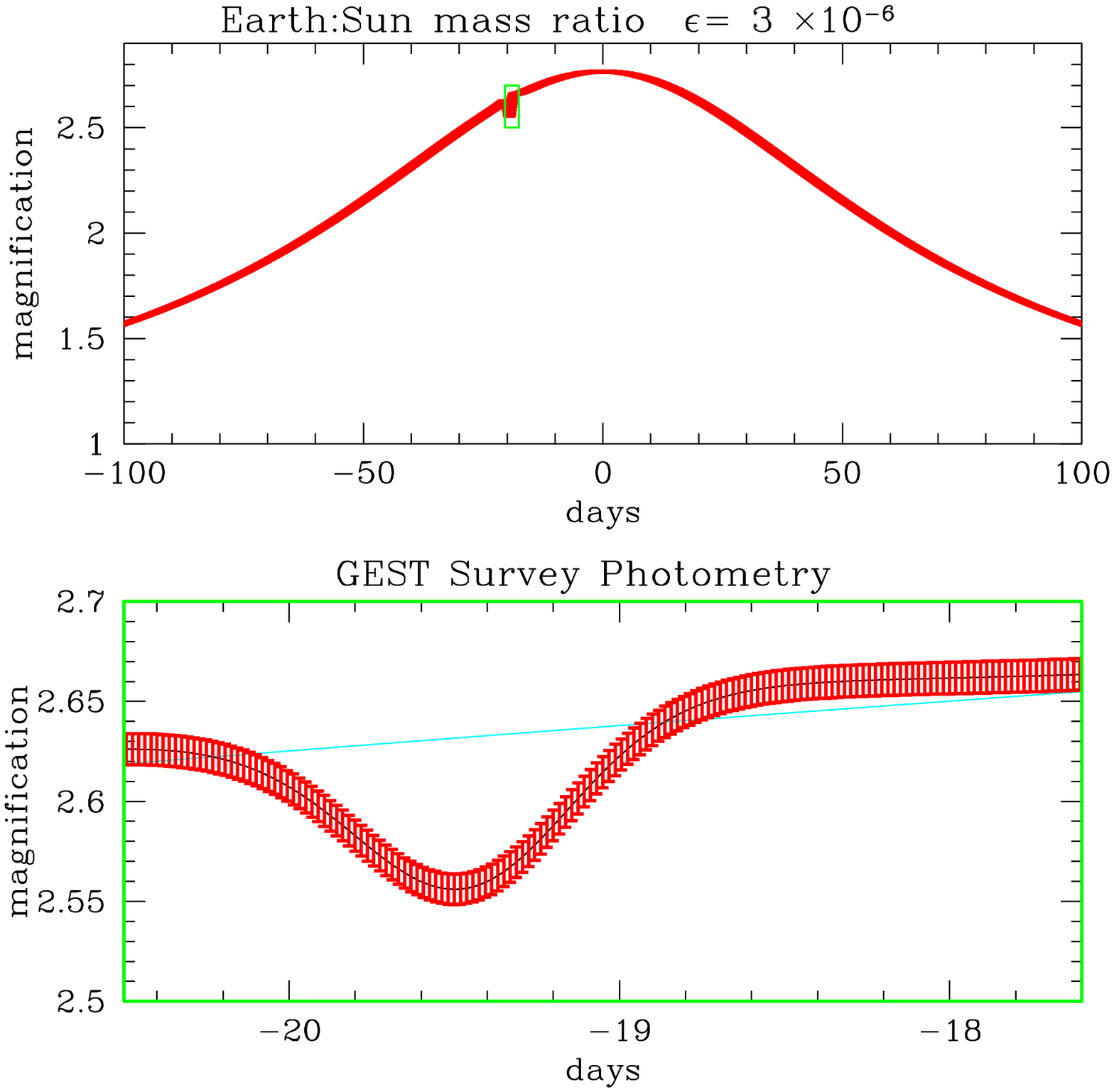}{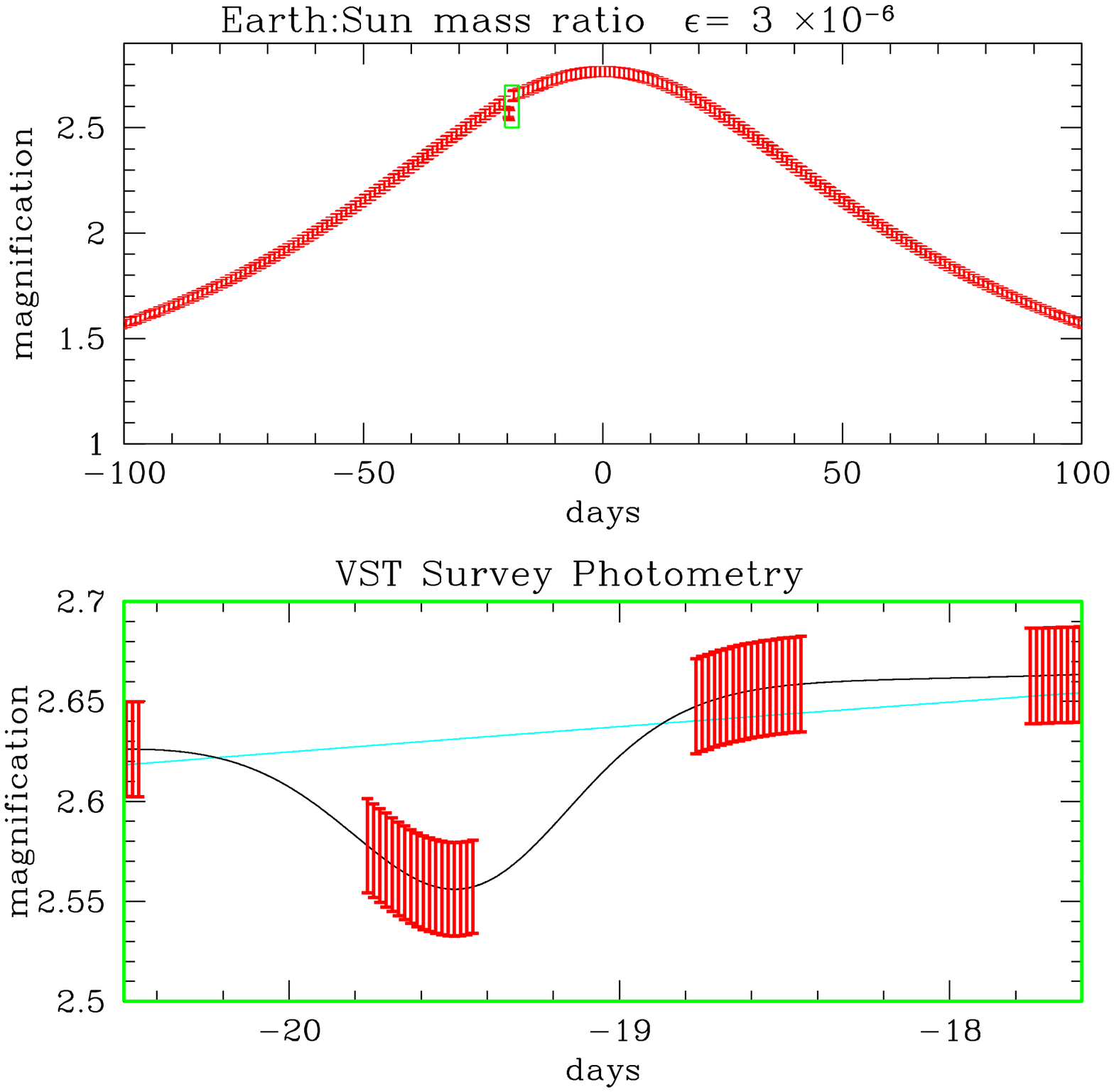}
\caption{A comparison of the same planetary microlensing event as seen 
by the GEST and VST simulations. The planet has the same mass fraction as the
Earth, $\epsilon = 3\times 10^{-6}$ with a separation of $a = 0.97$
Einstein ring radii. GEST detects the planet with a signal of 
$\Delta\chi^2 = 3100$, while The VST survey signal is $\Delta\chi^2 = 160$.
The VST survey data are probably insufficient to determine the planetary
parameters.
\label{fig-lcmarsorb15}}
\end{figure}

\begin{figure}
\plottwo{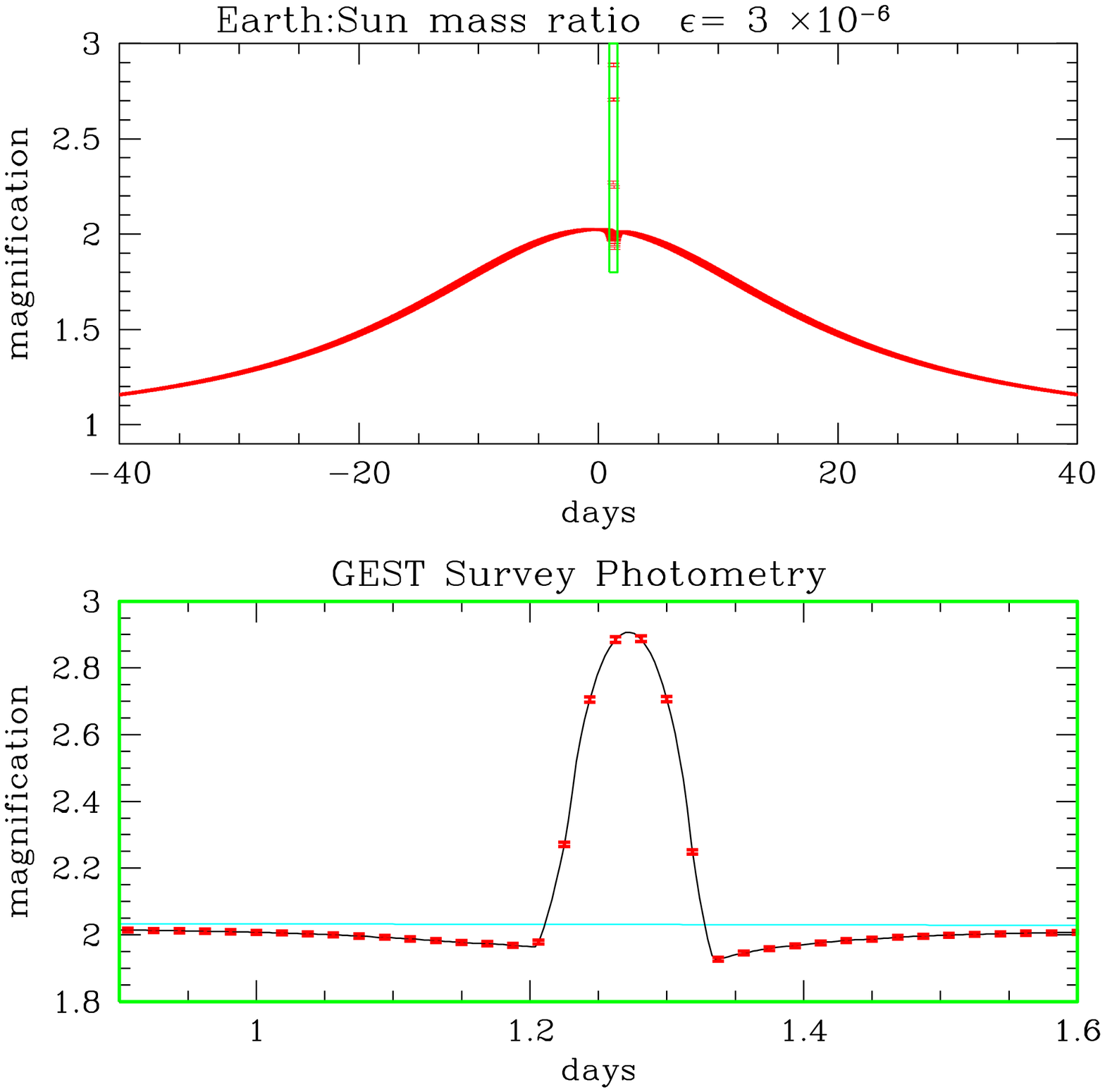}{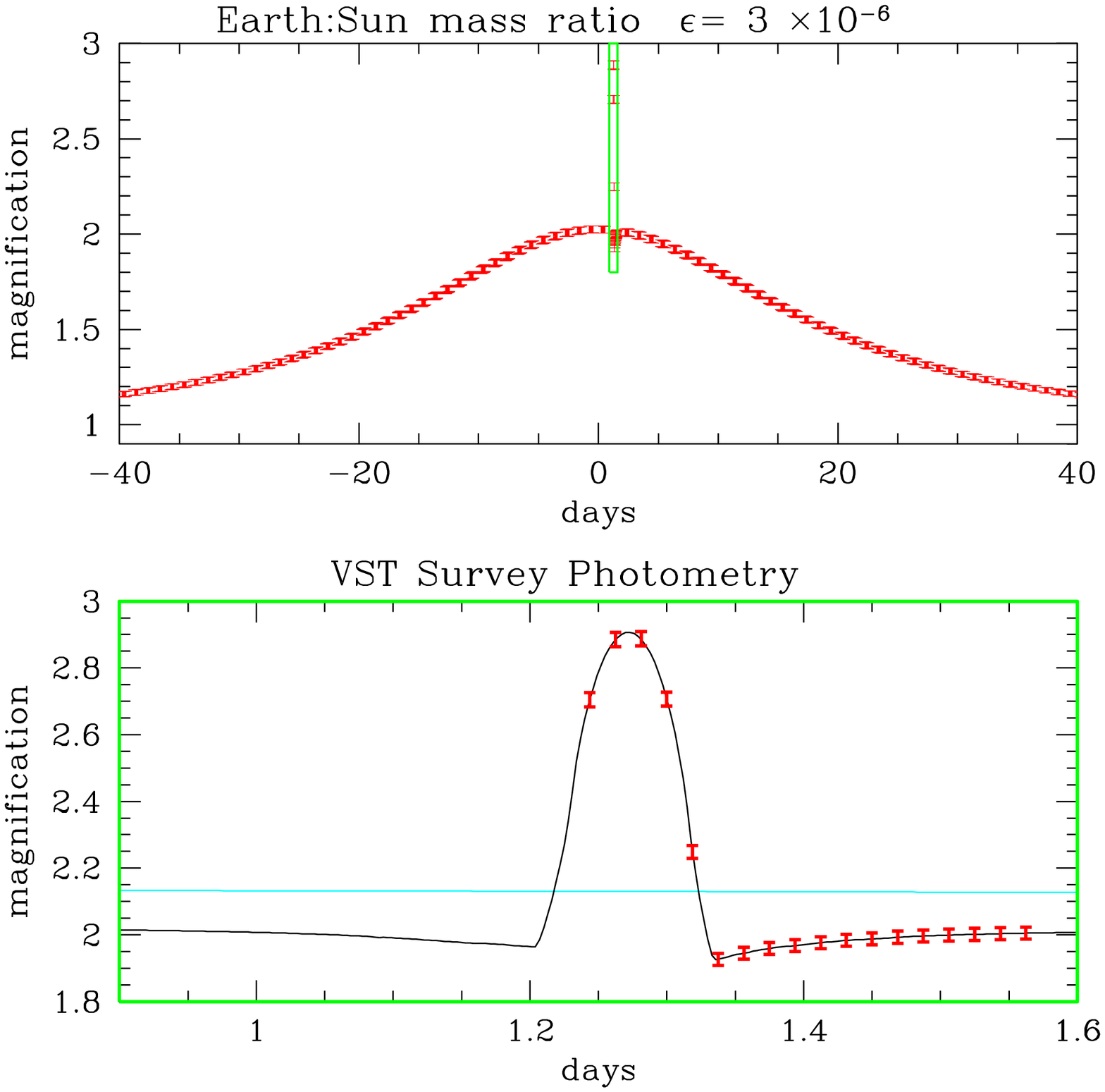}
\caption{The most significant detection ``fully sampled" Earth mass
fraction planet detection from the simulated VST survey is compared with 
the same event as seen in the GEST simulation. The separation is $a=1.31$
and the detection signal strengths are $\Delta\chi^2 = 39000$ and
$\Delta\chi^2 = 5400$ for the GEST and VST simulations, respectively.
Note that despite passing our criteria, the VST data does not fully 
sample the planetary deviation.
\label{fig-lcsatorb22}}
\end{figure}
 
\begin{figure}
\plottwo{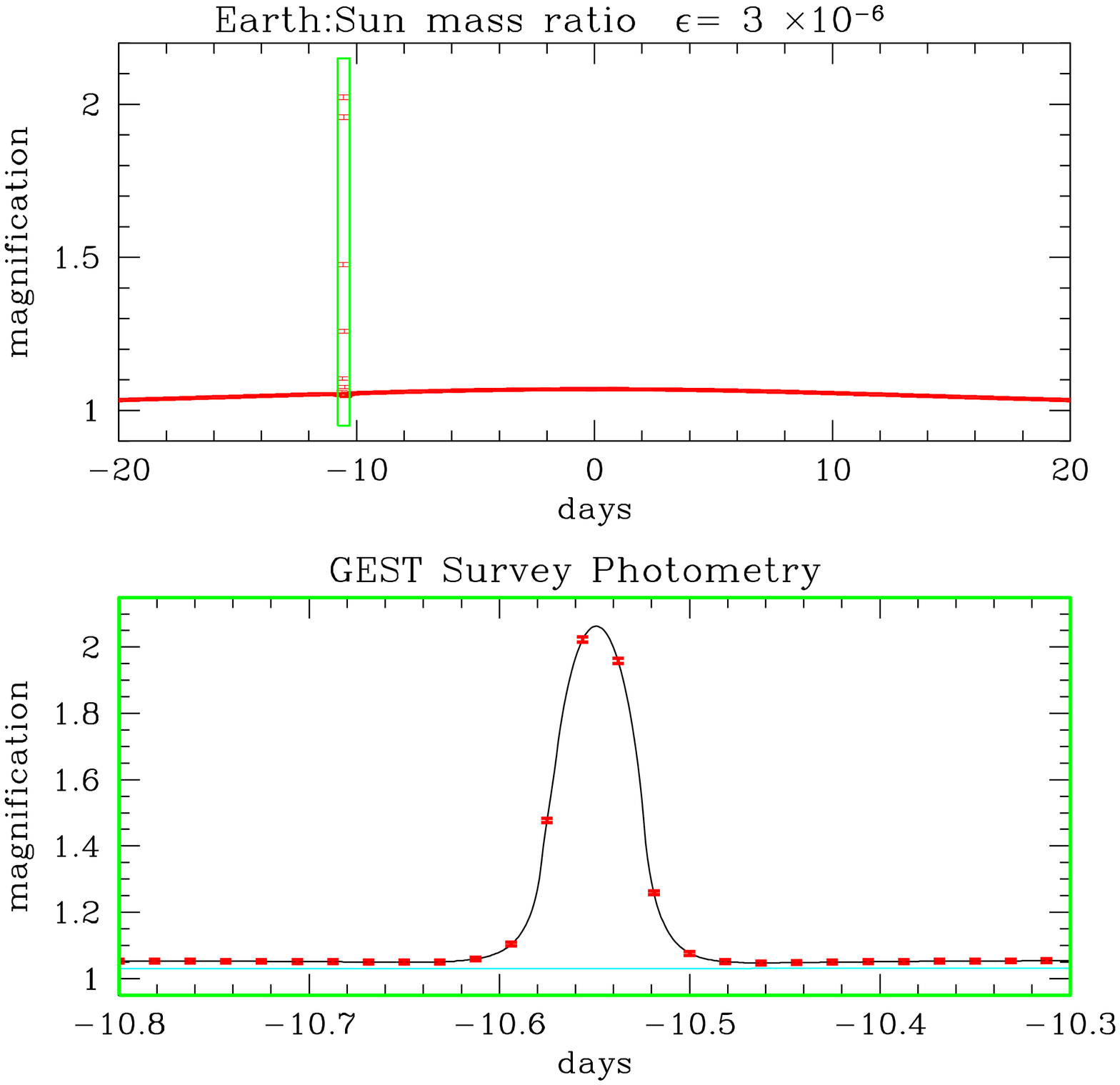}{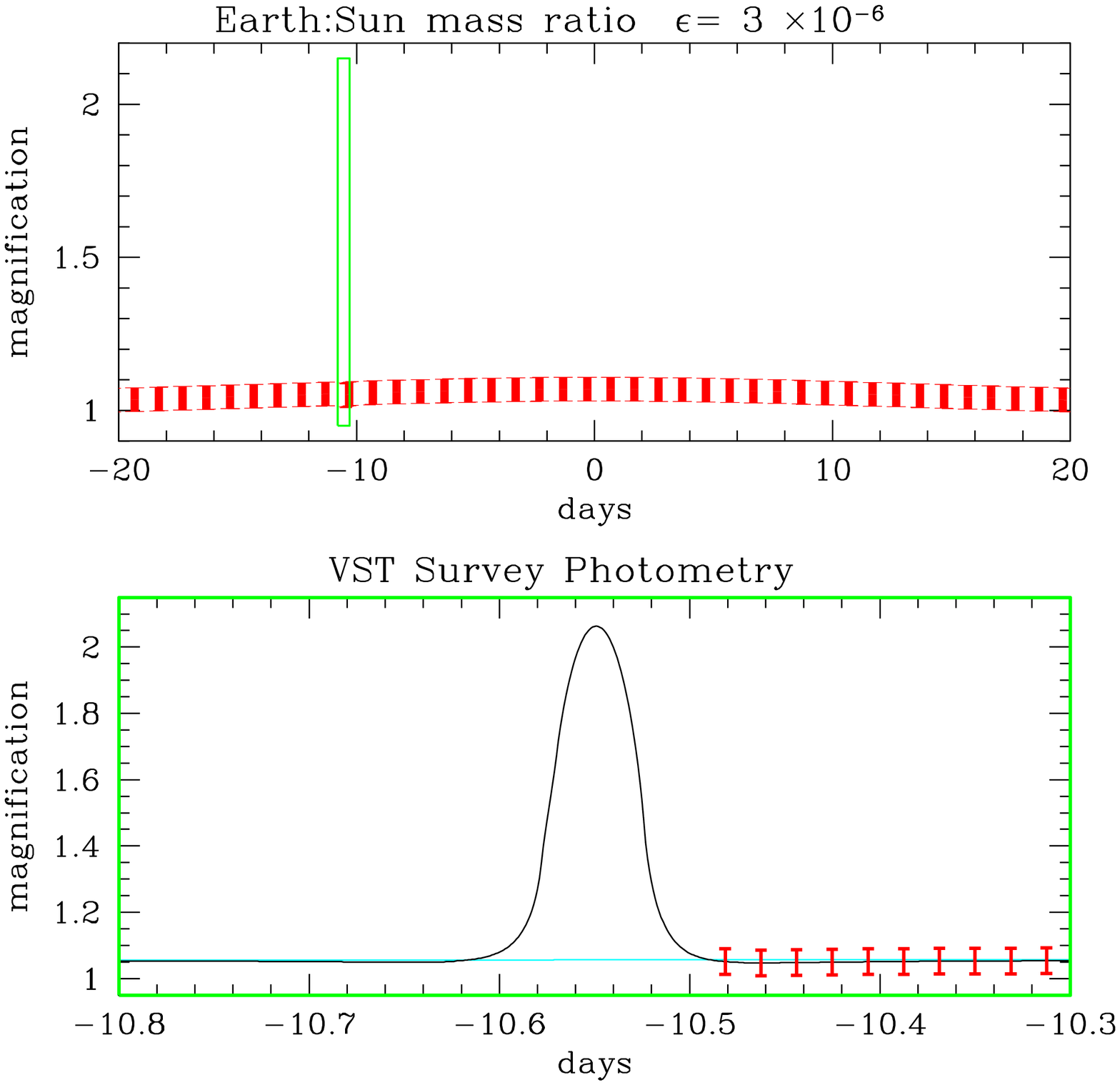}
\caption{A large amplitude planetary signal due to a 
$\epsilon = 3\times 10^{-6}$, $a=2.47$ planet is seen in a low magnification
stellar microlensing event for the GEST simulation, but both lensing event
is entirely missed in the VST simulation.
\label{fig-lcsatorb14}}
\end{figure}
 
In addition to detecting the planetary perturbation to the microlensing
lightcurve, it is also important to determine the characteristics of the 
planet that has been detected. Microlensing generally allows the determination
of the planetary mass fraction, $\epsilon$, and the transverse separation of 
the planet from the lens star in units of the Einstein radius which is
typically about $3(M_{\rm star}/\msun)$AU. Gaudi and Gould (1997) have shown
that these parameters can be accurately determined if the lightcurve
deviations are well sampled. For caustic crossing planetary microlensing events
which comprise a large fraction of the low mass planet detections, it is also 
possible to determine the planetary mass to about a factor of 2 or 3.
 
The gravitational microlensing planet search technique has previously
been been considered for ground based observations (Peale 1997; Sackett 1997;
Rhie et al 2000; Albrow et al 2000),
but ground based surveys face some difficulties due to the requirement
of continuous lightcurve monitoring. This can be accomplished with a network
of microlensing follow-up telescopes spanning the globe at southern latitudes,
but this requires that the survey rely upon observing sites that often have
poor weather or seeing conditions. Sackett (1997) has argued that a dedicated
2.5m telescope like the VST
with a wide field camera at an excellent site like Paranal 
could efficiently search for low mass planets where the deviations are 
expected to last only a few hours. We have done a simulation of such a survey
(optimistically) assuming observations in consistent 0.7" seeing for 8 hours
every night for 3 bulge seasons, and we find that low mass planets are
not easily detected by such a ground based survey because the highest
signal-to-noise events generally last longer than a few hours. If we demand
that more than 90\% of the $\Delta\chi^2$ signal occur in the 8 hour
observing window for an event to have measurable parameters, then we find
that this ``VST" survey is 30-50 times less sensitive to low mass planets
than GEST as shown in Figure 2. Also, even the ``best" planet detection
in the VST survey misses a significant part of the planetary deviation
as seen in Figure 6. A study by Peale (1997) of a ground based microlensing
planet search program with follow-up telescopes in Chile, Australia, and
South Africa, also only manages a handful of low mass planet detections
after eight years of observations and is not sensitive to isolated or
Mars mass planets. So, the proposed GEST mission is $\simgt 50$ times
more sensitive to low mass planets than both types of propsed ground based 
surveys.

\end{document}